\documentstyle[aps,epsf,preprint,amssymb]{revtex}

\def\beq{\begin{equation}}
\def\eeq{\end{equation}}
\def\beqa{\begin{eqnarray}}
\def\eeqa{\end{eqnarray}}

\def\frac#1/#2{\leavevmode\kern.1em
   \raise.5ex\hbox{\the\scriptfont0 #1}\kern-.1em
   /\kern-.15em\lower.25ex\hbox{\the\scriptfont0 #2}}

\begin{document}
\title{GEOMETRIC PHASE IN SU(N) INTERFEROMETRY}
\author{Hubert de Guise\,\footnote{now at Facult\'e Saint-Jean, University 
of Alberta, 8406 rue Marie-Anne Gaboury, Edmonton, T6C 4G9, Canada}}
\address{C.R.M., Universit\'e de Montr\'eal, C.P. 6128-A, Montr\'eal, H3C 3J7,
Canada}
\author{Barry C. Sanders, Stephen D. Bartlett and Weiping Zhang}     
\address{Department of Physics and Center for Lasers and its applications, 
Macquarie University, Sydney, N.S.W. 2109, Australia}
\maketitle

\begin{abstract}
An interferometric scheme to study Abelian geometric phase shift 
over the manifold SU(N)/SU(N-1) is presented.
\end{abstract}

\section{Introduction}     

The purpose of this contribution is twofold: to review how an SU(N) 
transformation can be experimentally realized using optical elements, and
to show how such an experimental realization can be used to investigate
the cyclic evolution of a state over the manifold SU(N)/U(N-1).
The bulk of the results will be presented explicitly for SU(3)
(see \cite{ourpaper} for further details) and SU(4),
although it will become clear that the method can be applied to any SU(N). 

Recall that cyclic evolution of a wave function yields the original state plus
a phase shift, and this phase shift is a sum of a dynamical phase $\varphi_d$
and a geometric (or topological, or quantal, or Berry) phase $\varphi_g$ 
shift\cite{Ber84,Sim83}.  
The geometric phase shift is important
not just for quantum systems but also for all of wave physics.
Thus far, controlled geometric--phase experiments, both realized and proposed,
have been exclusively concerned with the abelian geometric phase 
arising in the evolution of U(1)--invariant states 
\cite{Ber84,Tom86,Kwi91,Sim93}

Here, we generalize the above results to an Abelian geometric phase
which arises from geodesic transformations of U(N-1)--invariant states in 
SU(N)/U(N-1) space. The scheme employs a sequence of optical element, 
henceforth called SU(N) elements because they perform transformations 
described by an SU(N) matrix, arranged so  that the net result of the sequence
cyclically evolves an 
initial state back to itself up to a phase.  It will be seen 
that the decomposition of 
an SU(N) transformation into a product of appropriate SU(2) 
subgroup transformations is the
prescription to construct each SU(N) element as a sequence of SU(2) elements.

It is important to distinguish the evolution of states in the 
geometric space SU(N)/U(N-1) from the transformations of the optical beam 
as it  progresses through the interferometer.  It is possible to set up the
experiment so as to eliminate 
the dynamical phase associated with these optical transformations, thus making
the dynamical phase irrelevant for our purpose. The 
cyclic evolution described here occurs
in the geometric space, and the geometric phase of interest is related
to this evolution.

\section{SU(N) optical elements}

Consider an optical element which mixes two input beams.  It is, formally,
a black box which performs some transformation, as the output is not
the same as the input.  We are here interested by optical elements which
mix the input beams in a linear way, {\it i.e.} the output is a linear
combination of the inputs.   Furthermore, we will assume that the optical
element is passive, {\it i.e.} it does not globally create or annihilate
photons.  

The optical elements the enter in the construction of SU(N) device are 
beam splitters, mirrors and phase shifters.   A phase shifter is essentially a
slab of material which
increases the optical path lenght of one beam relative to the other.  A beam
splitter is a partially--silvered mirror which lets photons through with
some probability. 

Provided that losses can be ignored, each of these optical elements
can be associated with an SU(2) unitary transformation\cite{Yur86,Cam89}.
It is therefore advantageous to factorize each SU(N)
transformation into a product of SU(2)$_{ij}$ subgroup transformations 
mixing fields $i$ and $j$.

An optical element mixing two fields is associated with an SU(2) transformation
in the following way.
Suppose that {\it one} photon enters the black box.  We may assume
that it will 
enter the optical system either via beam one or beam two.  Thus the
Hilbert space of input states is two--dimensional.  As the transformation is
linear, the set of all possible output states will also be a two--dimensional
space.  Clearly this conclusion does not change if the input state is 
a general state 
$(\alpha,\beta)^t$, where the photon has probability $\vert \alpha \vert^2$
of being in beam 1 and probability $\vert \beta \vert^2$ of being in beam 2 
($\alpha$, $\beta$ are complex numbers).

Suppose now that {\it two} photons enter the black box.  Then, we can have
one of three possibilities: two photons enter in beam 1, one photon enters in
each beam, or both photons enter in beam 2.  In this case, the Hilbert space
of states is three--dimensional.

Continuing in this way, and using the fact that the input photons are 
indistinguishable, one rapidly works out that, in a system containing
$\lambda$ photons, the relevant Hilbert space is of dimension $\lambda+1$.

The conservation of photon number leads to the following constraint on
the possible form of the linear transformation.  Consider first the case of
a single photon.  The optical transformation 
\begin{equation}
T=
\left(
\begin{array}{cc}
a&b\\
c&d\end{array}\right)\, ,
\end{equation}
with $a, b, c, d$ complex numbers, transforms a general state 
$(\alpha_{in},\beta_{in})^t$ into the output state 
$(\alpha_{out},\beta_{out})^t$  such that
\begin{equation}
\left(\begin{array}{cc}\alpha_{out}\\ \beta_{out}\end{array}\right)
=
\left(
\begin{array}{cc}
a&b\\
c&d\end{array}\right)
\,
\left(\begin{array}{cc}\alpha_{in}\\ \beta_{in} \end{array}\right)\, .
\end{equation}
Taking the transpose complex conjugate of that to find 
$(\alpha_{out}^*,\beta_{out}^*)$, multiplying from the right by 
$(\alpha_{out},\beta_{out})^t$, we find that, if the number of photon
(=1) is to be conserved for any input state, 
$\vert \alpha_{out}\vert^2 +\vert \beta_{out}\vert^2 =
\vert \alpha_{in}\vert^2 +\vert \beta_{in}\vert^2=1$ implies that
$T^{\dagger}\cdot T = 1$, the unit matrix.  Thus, $T$ is a $2\times 2$ unitary 
transformation.  Because we are only interested in the relative phase 
between the beams, the determinant of $T$ can chosen without loss generality to
be +1, so that $T$ is an SU(2) matrix.

If the black box performs a transformation $T$ that is an SU(2) transformation
when there is a single photon in the system, it must also perform an SU(2)
transformation when there are $\lambda +1$ photons in the system: the 
transformtation effected by the black box cannot depend on the number of
photons in the system (at least not in the regimes that we are considering).
Thus, in a system of two photons, where state space is three--dimensional,
$T$ will be $3\times 3$ representation of the relevant SU(2) transformation.
In a system containing $\lambda$ photon, $T$ will be an SU(2) matrix of
dimension $(\lambda+1)\times(\lambda+1)$\cite{Yur86,Cam89}.

It is well known that an SU(2) transformation can be factored into a 
product of three subtransformations:
\begin{equation}
\begin{array}{ccccc}
R_z(\alpha)&\cdot& R_y(\beta) & \cdot & R_z(\gamma)\\
\left(\begin{array}{cc}e^{i\alpha}&\\&e^{-i\alpha}\end{array}\right)
&&
\left(\begin{array}{cc}\cos\beta&-\sin\beta\\ 
\sin\beta&\cos\beta \end{array}\right)&&
\left(\begin{array}{cc}e^{i\gamma}&\\&e^{-i\gamma}\end{array}\right)
\end{array}
\end{equation}

This factorization is a prescription on how to construct the SU(2) device:
a slab of material is inserted in one beamline so as to create a relative
phase shift of $e^{2i\gamma}$, a partially silvered mirror which lets
$\cos^2\beta$ photons from beam 1 through is then inserted, and another
phase shifter completes the design.

In an SU(3) interferometer, an general SU(3) matrix is decomposed into a 
product of three SU(2) matrices \cite{Row99}:
\begin{equation}
\left(\begin{array}{ccc}
1& 0     &0\\
0& a_1   &b_1\\
0& -b_1^*&a_1^*\end{array}
\right)
\cdot 
\left(\begin{array}{ccc}
e^{i\alpha}\cos t& -\sin t           & 0\\
\sin t           & e^{-i\alpha}\cos t& 0\\
0                &0                  &1\end{array}
\right)\cdot
\left(\begin{array}{ccc}
1& 0     &0\\
0& a_2   &b_2\\
0& -b_2^*&a_2^*\end{array}\right)\, ,
\end{equation}
where $\vert a_i\vert ^2 + \vert b_i\vert^2 =1$.

This factorization, symbolically written $R^1_{23}\cdot R_{12}\cdot R^2_{23}$,
is a {\it de facto} prescription on how to build the SU(3)
device: fields $2$ and $3$ are mixed followed by a mixing of the output
field~$2$ with the field in channel~$1$, and, finally,
the output field~$2$ is mixed with field~$3$.

As it is possible to factorize an SU(N) matrix in terms of SU(2) submatrices
\cite{Rec94},
the process of constructing a general SU(N) device is perfectly obvious and
follows the lines illustrated explicitly for the SU(3) device.

For instance, the appropriate factorization of an SU(4) matrix is a product
of the type 
\begin{equation}
R^1_{23}\cdot R^1_{34}\cdot R^2_{23}\cdot R_{12}
\cdot R^3_{23}\cdot R_{34}^2\cdot R_{23}^4\, ,
\end{equation}
where $R^i_{k\ell}$ is an SU(2) matrix mixing fields $k$ and $\ell$.

The factorization of an SU(N) matrix into SU(2) subgroup matrices is not 
unique, and the number of SU(2) elements required to construct an SU(N)
device can vary according to the parametrization: an estimate of the number
of SU(2) devices required to construct an SU(N) element was given in 
\cite{torma}.  Finally, we mention that SU(3) and 
SU(4) devices have been constructed but with an aim to study non-classical 
statistics \cite{mattle}.
 
\section{Geodesic evolutions}

The total phase $\varphi$ acquired by a state during a generic cyclic 
evolution is the sum of $\varphi_d+\varphi_g$. 

A special type of evolution is the geodesic evolution\cite{Chy88};
by transforming the output state along geodesic paths in
the geometric space, the geometric phase shift along each path is zero.

An essential property of geodesic evolutions is that they are not transitive:
the product of two such evolutions is not necessary another geodesic evolution.
This is most easily illustrated by drawing three points on a plane at random.
It is well known that the geodesic on a plane is a straight line.  Let 
$|1\rangle\, ,|2\rangle$ and $|3\rangle$ denote the three points.  Then it is
clear that, even if $R_{12}$ is the straight (geodesic) line that connects 
$|1\rangle$ and  $|2\rangle$, and even if $R_{23}$ is the straight line that 
connect $|2\rangle $ and $|3\rangle$, the combined segment 
$R_{23}\cdot R_{12}$ is {\it not} a geodesic between $|1\rangle $ and
$|3\rangle$.  

This property makes it possible to construct a cyclic evolution from a sequence
of geodesic legs: the geometric phase acquired during the circuit is then
a global property of the entire circuit.  

For definiteness, let us consider SU(3).  There, 
the evolution of the state~$\psi^{(1)}$ 
to the state $\psi^{(4)} = e^{i\varphi_g}\psi^{(1)}$
via 3 geodesic paths in the geometric space can be described by 
3 one--parameter SU(3) group elements $\{ U^g_k(s_k);\ k=1,2,3 \}$,
with $s_k$ an evolution parameter.
These transformations satisfy the conditions that $U^g_k(0)$ is the 
identity element and
\begin{equation} 
  \label{U_k}
  U^g_k(s_k^0) \psi^{(k)} = \psi^{(k+1)} , \quad k =  1,2,3 \, ,
\end{equation}
for some fixed end values $\{ s_k^0 \}$ of the evolution
parameters. 
We consider evolutions $U_k^g(s_k)$ of the form
\begin{equation}
  \label{eq:DecomposeU}
  U_k^g(s_k)= V_k\cdot R_{s_k}\cdot V_k^{-1}\, ,
\end{equation}
with $V_k$ an element of SU(3) satisfying $\langle \psi^{(k)} | U^g_k(s_k)
| \psi^{(k)} \rangle$ real and positive, and
\begin{equation} 
  \label{eq:rt}
  R_{s_k}       \equiv \left( \begin{array}{ccc}
                \cos s_k & -\sin s_k & 0 \\
                \sin s_k & \cos s_k & 0 \\
                0&0&1\end{array}\right)\, .
\end{equation}
The form of the one--parameter subgroup $R_{s_k}$ with real
entries was guided by the
definition of a geodesic curve between two states $\psi^{(k)}$ and
$\psi^{(k+1)}$, which can be written in the form \cite{Arv97}
\begin{equation}
  \label{eq:geocurve}
  \psi(s_k) = \psi^{(k)}\cos s_k 
  +{ \left(\psi^{(k+1)}-\psi^{(k)}
  \langle \psi^{(k+1)}\vert \psi^{(k)}\rangle \right) \over 
  \sqrt{1- \langle \psi^{(k+1)}\vert \psi^{(k)}\rangle^2 }}
  \sin s_k \, ,
\end{equation}
with $0\le s_k \le s_k^0 = \arccos \langle \psi^{(k+1)}\vert 
\psi^{(k)}\rangle$.   As it is always possible to
choose unit vectors $\psi^{(k)}$ such that 
$\langle \psi^{(k+1)}\vert \psi^{(k)}\rangle$ is real and positive,
it is straightforward to show that any $U^g_k(s_k)$
of the form given by Eq.\ (\ref{eq:DecomposeU}) satisfying $\langle 
\psi^{(k+1)}\vert \psi^{(k)}\rangle$ real and positive gives evolution
along a geodesic curve in SU(3)/U(2).

The form of the geodesic evolution makes it easy to obtain its physical
interpretation.  The transformation
$R_{s_k}$ is a transformation of appropriate length along some
reference geodesic (some generalized Greenwich meridian on SU(3)/U(2)).  
The transformation
$V_k$ is a principal axis transformation which correctly orients the 
reference geodesic so that it passes through $|\psi^{k}\rangle$ and
$|\psi^{k+1}\rangle$.  $V_k$ therefore depends on the initial 
and final states.  

The three states in SU(3)/U(2) must 
be chosen in a sufficiently general way to ensure
that they can represent any triangle in SU(3)/U(2) \cite{Arv97}.  
Since the latter is of dimension $4$, there are $4$ free parameters to be 
chosen.  The first state can be chosen, WLOG, to be the ``north pole'' state. 
Again WLOG, the second state can always be chosen to lie along the reference 
geodesic some distance away from the initial state.  The last state must 
therefore contain the remaining $3$ parameters.  In short, the vertices of a 
geodesic triangle in SU(3)/U(2) can, in general, be chosen as 
\begin{eqnarray}
  \label{vertices}
  \psi^{(1)} &=& \left( \begin{array}{c} 1 \\ 0 \\ 0 \end{array} \right)
        = e^{-i\varphi_g} \psi^{(4)} , \quad
  \psi^{(2)} = \left( \begin{array}{c} \cos s_1^0 \\ \sin s_1^0 \\ 0 
  \end{array} \right) ,
                \nonumber \\
  \psi^{(3)} &=&  \left( \begin{array}{c}
        \cos s_1^0 \cos s_2^0 - e^{i\alpha} \sin s_1^0 \sin s_2^0
                \cos \beta \\
        \sin s_1^0 \cos s_2^0 + e^{i\alpha} \cos s_1^0 \sin s_2^0
                \cos \beta \\
        \sin \beta \sin s_2^0
        \end{array} \right) ,
\end{eqnarray}
with $s_1^0$, $s_2^0$, $\alpha$ and $\beta$ arbitrary. 

Since $|\psi^4\rangle = e^{i\varphi_g}|\psi^1\rangle$, the geometric phase
for the complete 
circuit is extracted from the overlap real positive overlap
$\langle \psi^{(3)}\vert \psi^{(1)}\rangle$.  This works out immediately to 
\begin{equation}
\varphi_g = 
\hbox{\rm arg}(\cos s_1^0 \cos s_2^0 - e^{-i\alpha} \sin s_1^0 \sin s_2^0
                \cos \beta\,) .\label{berryphase}
\end{equation}

The generalization to SU(4) is immediate: the form of Eq.(\ref{eq:DecomposeU})
remains the same, but the matrix of Eq.(\ref{eq:rt}) is augmented to a
$4\times 4$ matrix:
\begin{equation}
R_{s_k}=\left(\begin{array}{cccc}
\cos s_k&-\sin s_k&0&0\\
\sin s_k& \cos s_k&0&0\\
0       & 0       &1&0\\
0       & 0       &0&1\end{array}\right)\, .
\end{equation}

The condition of Eq.(\ref{eq:geocurve}) remains.  As we have argued, the
first two vertices of the geodesic triangle remain unchanged, but the last
vertex now depends on the $6$ parameters of SU(4)/U(3):
\begin{equation}
{\renewcommand{\arraycolsep}{0.8pt}
\begin{array}{ll}
\psi^{(1)} &= \left( \begin{array}{c} 1 \\ 0 \\ 0 \\ 0\end{array} \right)
        = e^{-i\varphi_g} \psi^{(4)} , \qquad
  \psi^{(2)} = \left( \begin{array}{c} \cos s_1^0 \\ \sin s_1^0 \\ 0 \\0
  \end{array} \right) , \\
&\\
  \psi^{(3)} &=  \left( \begin{array}{c}
        \cos s_1^0 \cos s_2^0 -e^{i\alpha} \sin_1^0 \sin s_2^0 
\left( -\cos\beta_1 \cos\beta_2 +\sin\beta_1\sin\beta_2\cos\beta_3\right)\\
        \sin s_1^0 \cos s_2^0 + e^{i\alpha} \cos s_1^0 \sin s_2^0
\left(\cos\beta_1\cos\beta_2-\sin\beta_1\sin\beta_2\cos\beta_3\right)\\
\left(\cos\beta_1\sin \beta_2+\sin\beta_1\cos\beta_2\cos\beta_3\right)
\sin s_2^0\\
\sin s_2^0 \sin\beta_1 \sin\beta_3
        \end{array} \right) ,
\end{array}}
\end{equation}
a form which obviously reduces to the SU(3) case if $\beta_3 =0$.
For SU(4)/U(3), the Berry phase is again related to the inner product
of $\langle \psi^{(1)}\vert \psi^{(3)} \rangle$ through
$\vert \psi^{(4)}\rangle = e^{i\varphi_g}\vert \psi^{(1)}\rangle$ and 
can be seen to depend on the required number of parameters.

\section{Geometric phase in SU(N) interferometry}

An optical SU(N) transformation can be realized by a N--channel
optical inter\-fe\-ro\-meter\cite{Rec94}. 

The SU(2)$_{12}$ matrix~$R_s$ in Eq.\ (\ref{eq:rt}) is a special case 
of the generalized lossless beam splitter
transformation for mixing channels~$1$ and~$2$.
More generally a beam splitter can be described by a unitary transformation 
between two channels\cite{Cam89}.
For example, a general SU(2)$_{23}$ beam splitter transformation for 
mixing channels~$2$ and $3$ is of the form
\begin{equation}
  R_{23}(\phi_{\rm t},\theta,\phi_{\rm r})
        = \left(\begin{array}{ccc}
        1&0&0\\
        0&e^{i\phi_{\rm t}}\cos\theta&-e^{-i\phi_{\rm r}}\sin\theta\\
        0&e^{i\phi_{\rm r}}\sin\theta
                &e^{-i\phi_{\rm t}}\cos\theta\end{array}
        \right) \, ,
\end{equation}
with~$\phi_{\rm t}$ and $\phi_{\rm r}$
the transmitted and reflected phase--shift parameters, 
respectively, and $\cos^2\theta$ the beam splitter transmission.
A generalized beam splitter can be realized as a combination of
phase shifters and 50/50 beam splitters in a Mach--Zehnder 
interferometer configuration.

The goal of the following is to construct an SU(3) optical
transformations in terms of SU(2) elements which realize the geodesic
evolution in the geometric space by appropriately adjusting
parameters of the interferometer.

It will be convenient to write $\psi^{(3)}$ in 
Eq.(\ref{vertices}) as 
$(e^{i\xi}\cos\eta, e^{i(\xi+\chi)} \sin\eta\cos\tau, \sin\eta\sin\tau)^t$, 
where $\xi$, $\eta$, $\tau$ and $\chi$
are functions of $s_1^0$, $s_2^0$, $\alpha$ and $\beta$, the parameters of
$\psi^{(3)}$ in Eq.\ (\ref{vertices}).
Following our factorization scheme, the geodesic evolution operators
$U^g_k(s_k)$,
connecting $\psi^{(k)}$ to $\psi^{(k+1)}$,
can be expressed as
\begin{eqnarray}
  \label{evolutions}
  U^g_1(s_1)
        &=& R_{s_1}\, ,\nonumber \\
  U^g_2(s_2)
        &=& R_{s_1^0}\cdot R_{23}(\alpha,\beta,0)\cdot R_{s_2}  
        \cdot R^{-1}_{23}(\alpha,\beta,0)\cdot R_{-s_1^0}\, 
        ,\nonumber \\
  U^g_3(s_3)&=&
        R_{23}(\chi,\tau,\xi) \cdot R_{-s_3} \cdot 
        R^{-1}_{23}(\chi,\tau,\xi)
        \, ,
\end{eqnarray}
with $R_s$ given by Eq.\ (\ref{eq:rt}), the parameters $s_k$ ranging from 
$0\leq s_k \leq s_k^0$, and $s_3^0 = \eta$.  Note that $s_3^0$ and, in 
fact, all the parameters of $U^g_3(s_3)$ are fixed by the requirement
that $\psi^{(4)} = e^{i\varphi_g}\psi^{(1)}.$
Also note that, for each $k$, $U^g_k(0)$ is
the identity in SU(3) and $U^g_k(s_k^0)\psi^{(k)} = \psi^{(k+1)}$
as required.  Once it is observed that
$\langle \, \psi^{(k+1)}\vert \ \psi^{(k)}\, \rangle = \cos s_k^0$,
it is trivial to verify that each evolution satisfies Eq.\ 
(\ref{eq:geocurve}) and is therefore geodesic.

The geometric phase for the cyclic evolution $\psi^{(1)} \to 
\psi^{(4)}$ is given explicitly by Eq.(\ref{berryphase}).
This phase depends on four free parameters in the experimental
scheme: $s_1^0$, $s_2^0$, $\alpha$ and $\beta$, which describe a general
geodesic triangle in SU(3)/U(2).

\begin{figure}[h]
\begin{center}
   \leavevmode
     \epsfysize=6.0cm  \epsfbox{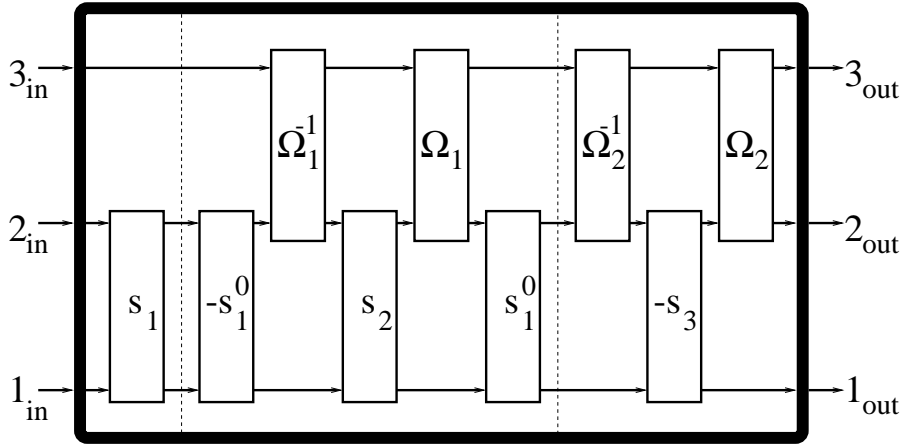}
        \caption{
The SU(3) interferometer is depicted, with three input ports
$1_{\rm in}$, $2_{\rm in}$ and $3_{\rm in}$, and three output ports 
$1_{\rm out}$, $2_{\rm out}$ and $3_{\rm out}$.
There are nine beam splitter transformations
with parameters $s_1$, $s_2$, $s_3$, $\Omega_1 = (\alpha,\beta,0)$ and
$\Omega_2 = (\chi,\tau,-\xi)$.
For geodesic, cyclic evolution of the output state, 
only four parameters are independent.
}
\label{fig:config}
\end{center}
\end{figure}

The interferometer configuration for realizing the
necessary evolution about the geodesic triangle
is depicted in Fig.\ \ref{fig:config}.
This configuration consists of a sequence of SU(2)$_{ij}$ transformations,
and we use the shorthand notation
$\Omega_i \equiv (\alpha_i , \beta_i , \gamma_i )$
to designate the three parameters associated with the
generalized beam splitter.
The three--channel interferometer consists of a sequence of
nine SU(2)$_{ij}$ transformations.  The field enters port~$1_{\rm in}$,
and the vacuum state enters ports~$2_{\rm in}$ and~$3_{\rm in}$.

For SU(4), the first evolution is the same, but the second will depend on 
more parameters as the dimensionality of SU(4)/U(3) is larger than that
of SU(3)/U(2).  Briefly, we have:
\begin{eqnarray}
  \label{su4evolutions}
  U^g_1(s_1)
        &=& R_{s_1}\, ,\nonumber \\
  U^g_2(s_2)
        &=& R_{s_1^0}\cdot \tilde V_2 \cdot 
R_{s_2}  
        \cdot \tilde V_2^{-1}\cdot R_{-s_1^0}\, 
        ,\nonumber \\
  U^g_3(s_3)&=&
        V_3 \cdot R_{-s_3} \cdot 
        V_3^{-1}
        \, ,
\end{eqnarray}
where
\begin{equation}
\tilde V_2 = R_{23}(\alpha_1,\beta_2,0)\cdot R_{34}(\alpha_1,\beta_3,0)\cdot
R_{23}(0,\beta_1,0)\, ,
\end{equation}
and where $V_3$ is an SU(3) matrix of the form $R_{23}\cdot R_{34}\cdot 
R^{\prime}_{23}$ whose details are unimportant for our purposes.

Although SU(3) and SU(4) interferometry have been considered in some details,
the methods employed here can be extended to SU(N),
or N--channel, interfe\-ro\-me\-try\cite{Rec94}.
The schemes discussed above employing such a device
would produce and enable observation of the geometric phase shift
for geodesic transformations of states invariant under \mbox{U(N-1)} 
subgroups of \mbox{SU(N)} states in the \mbox{2(N-1)}--dimensional
coset space \mbox{SU(N)/U(N-1)}.

\section{Discussion and Conclusion}

This contribution is a summary of recent theoretical work on the 
possibility of measuring Berry phases using optical elements.  The 
scheme depends on the optical realization of SU(N) transformations in the
optical domain; this is possible because the Lie algebra su(n) can be
realized in terms of boson creation and destruction operators which have
immediate interpretation as photon field operators.  There also exists the
possibility of realizing Sp(2n,${\Bbb R}$) transformation using optical
elements \cite{Yur86}: 
the Lie algebra sp(2n,${\Bbb R}$) also has a realization in
terms of boson operators.  The setup to measure Berry phase in an optical
experiment is interesting because it provides a very practical realization
of otherwise abstract ideas and allows one to do ``experimental differential
geometry'' over SU(N)/U(N-1).

This contribution has dealt exclusively with the optical realization of
Abelian Berry phase: even if states are invariant under U(N), two states are
equivalent if they differ by a U(1)--phase.  It is possible to enlarge the
equivalence class to obtain the so--called non-Abelian Berry phase 
\cite{Ana88}, which
has been studied in the context of degenerate states.  It is
possible to study the non-Abelian version of the results presented here
by using polarization: two states of different polarization are declared 
equivalent.  The larger equivalence class comes about because a rotation
of the polarization plane is an SU(2) transformation.  The 
experimental aspects of this remain, at the 
moment, unclear.  An optical experiment to  
measure an SU(2) phase would require optical devices which 
perform ``tunable'' polarization--dependent transformation.  The theoretical
aspects of this questions are currently under investigation.

This work has been supported by two Macquarie University Research Grants
and by an Australian Research Council Large Grant.
BCS appreciates valuable discussions with J.\ M. Dawes and A.\ Zeilinger,
and HdG acknowledges the support of Fonds F.C.A.R. of the Qu\'ebec Government.


\begin{thebibliography}{99}

\bibitem{ourpaper} B.\ C.\ Sanders, H.\ de\ Guise, S.\ D.\ 
Bartlett and Weiping Zhang, Phys.~Rev.~Lett. {\bf 86} 369 (2001)

\bibitem {Ber84} M.\ V.\ Berry, Proc.\ Roy.\ Soc.\ (Lond.)
        {\bf 392}, 45 (1984).
\bibitem {Sim83} 
        B.\ Simon, Phys.Rev.Lett {\bf 51}, 2167 (1983);
        F.\ Wilczek and A.\ Shapere, 
        {\em Geometric Phases in Physics},
        Advanced Series in Mathematical Physics - Vol. {\bf 5}
        (World Scientific, Singapore, 1989).
\bibitem {Tom86} A.\ Tomita and R.\ Y.\ Chiao, Phys.Rev.Lett {\bf 57}, 937 (1986);
        R.\ Y.\ Chiao, A.\ Antaramian, K.\ M.\ Ganga, H.\ Jiao,
        S.\ R.\ Wilkinson, and H.\ Nathel,
        Phys.Rev.Lett {\bf 60}, 1214 (1988);
        D.\ Suter, K.\ T.\ Mueller, and A.\ Pines,
        Phys.Rev.Lett {\bf 60}, 1218 (1988).
\bibitem {Kwi91} P.\ G.\ Kwiat and R.\ Y.\ Chiao, Phys.Rev.Lett {\bf 66}, 
588 (1991).
\bibitem {Sim93} 
        R.\ Simon and N.\ Mukunda, Phys.Rev.Lett {\bf 70}, 880 (1993).
\bibitem {Yur86} 
        B.\ Yurke, S.\ L.\ McCall, and J.\ R.\ Klauder,
        Phys.Rev.A {\bf 33}, 4033 (1986).
\bibitem {Cam89}
        R.\ A.\ Campos, B.\ E.\ A.\ Saleh and M.\ C.\ Teich,
        Phys.Rev.A {\bf 40}, 1371 (1989).
\bibitem {Row99} 
        D.\ J.\ Rowe, B.\ C.\ Sanders and H.\ de Guise,
        J.\ Math.\ Phys.\ {\bf 40}, 3604 (1999).
\bibitem {Rec94} M.\ Reck, A.\ Zeilinger, H.\ J.\ Bernstein, and P.\ Bertani, 
        Phys.Rev.Lett {\bf 73}, 58 (1994);
        B.\ C.\ Sanders, H.\ de Guise, D.\ J.\ Rowe and A.\ Mann, 
J.\ Phys.\ A: Math.\ Gen. {\bf 32}, 7791 (1999).

\bibitem {torma} P.\ T\"orm\"a, I.\ Jex and S.\ Stenholm, J.Mod.Opt. {\bf 43},
245 (1996).

\bibitem {mattle} K.\ Mattle, M.\ Michler, H.\ Weinfurter, A.\ Zeilinger, 
M.\ Zukowski, Appl.\ Phys.\ {\bf B 60}, S111 (1995).

\bibitem {Chy88}
        T.\ H.\ Chyba, L.\ J.\ Wang, L.\ Mandel, and R.\ Simon,
        Opt.\ Lett.\ {\bf 13}, 562 (1988);
        R.\ Simon, H.\ J.\ Kimble and E.\ C.\ G.\ Sudarshan,
        Phys.Rev.Lett {\bf 61}, 19 (1988).
\bibitem {Arv97}
        Arvind, K.\ S.\ Mallesh, and N.\ Mukunda, J.\ Phys.\ A: Math.\ Gen.
{\bf 30}, 2417 (1997)
\bibitem {Ana88} 
       J.\ Anandan, Phys.\ Lett.\ A {\bf 133}, 171 (1988), F. Wilczek and
A. Zee, Phys.\ Rev.\ Lett. {\bf 52}, 2111 (1984).
\end{thebibliography}
\end{document}